\documentclass[preprint,12pt,authoryear]{elsarticle}

\usepackage{amssymb}
\usepackage{longtable}
\usepackage{subcaption}
\usepackage{endfloat}
\usepackage{rotating, graphicx}
\usepackage{multirow}

\usepackage[usenames]{color}
\usepackage{colortbl}

\begin{document}

\begin{frontmatter}



\title{Spectroscopic Observations of the Comet 29P/ Schwassmann-Wachmann 1 at the SOAR Telescope}


\author[a,b]{Oleksandra V. Ivanova\corref{cor1}}
\address[a]{Astronomical  Institute  of  the  Slovak  Academy  of  Sciences,  SK-05960  Tatransk\'{a}  Lomnica,  Slovak Republic}
\address[b]{Main Astronomical Observatory, National Academy of Sciences of Ukraine, Goloseevo, Kyiv, 03680 Ukraine}
\cortext[cor1]{Corresponding author}
\ead{sandra@mao.kiev.ua}

\author[c]{Enos Picazzio}
\address[c]{Universidade de S$\tilde{a}$o Paulo, Instituto de Astronomia, Geofísica e Ci$\hat{e}$ncias Atmosf$\acute{e}$ricas, Departamento  de Astronomia, Rua do Mat$\tilde{a}$o, 1226; Cidade Universit$\acute{a}$ria, CEP 05508-900, S$\tilde{a}$o Paulo, SP, Brasil}

\author[d]{Igor V. Luk'yanyk}
\address[d]{Astronomical Observatory, Taras Shevchenko National University of Kyiv, Observatorna str. 3, Kyiv, 04053 Ukraine}

\author[e]{Oscar Cavichia}
\address[e]{Instituto de F\'isica e Qu\'imica, Universidade Federal de Itajub\'a, Av. BPS, 1303; Pinheirinho, CEP 37500-903, Itajub\'a, BRASIL}

\author[f,g]{Sergei M. Andrievsky}
\address[f]{Department of Astronomy and Astronomical Observatory, Odessa National University, Shevchenko Park, 65014, Odessa, Ukraine}
\address[g]{GEPI, Observatoire de Paris, PSL Research University, CNRS, Place Jules Janssen, 92195 Meudon, France}

\begin{abstract}
	We carried out photometric and spectroscopic observations of comet 29P/ Schwassmann-Wachmann 1 at the SOAR 4.1-meter telescope (Chile) on August 12, 2016. This paper presents the results of only spectroscopic analysis. The spectra revealed presence of CO$^+$ and N$_2^+$ emissions in the cometary coma at a distance of 5.9 AU from the Sun. The ratio [N$_2^+$]/[CO$^+$] within the projected slit seems to be 0.01. We have also estimated spectral gradient value for the comet.  
\end{abstract}


\end{frontmatter}


\section{Introduction}
%\label{}
Small bodies with orbits beyond Neptune orbit are of interest because they must have undergone minimal changes since the Solar System formation. Some comets show significant activity at distances far from the Sun. This is unusual since at the low equilibrium temperature ($<$140\,K) the significant physical activity caused by the water ice sublimation is not expected. At the distances greater than 3\,AU sublimation of CO and CO$_2$ ices may become the triggering mechanism producing the coma. Optical spectroscopy reveals the presence of the COþ ions and CN radicals in the coma of comet 29P/Schwassmann-Wachmann 1 (hereafter SW1) (\citealt{Cochran1991a}; \citealt{Cochran1980}; \citealt{Cochran1982}; \citealt{Cochran1991b}; \citealt{Cook2005}; \citealt{Larson1980}).\\
Current observational data (including Galileo space telescope data) indicate the high abundance of the metals (from astrophysical point of view, all elements heavier than helium) with respect to hydrogen in the giant planets atmospheres, and this value is much higher than that observed in the Sun. To explain this fact, the giant planets formation models require presence of accretion. The accreted material should contain the volatile components, such as neon, argon, krypton, xenon, which could not survive at the distance where giant planets were formed. Such a substance could be effectively delivered by the Oort Cloud and Kuiper Belt bodies, assuming that they were formed at temperatures below 30 K. This fact is supported by our observations of distant comets, after the emission line N$_2^+$ was detected in two comets, C/2002 VQ94 (LINEAR) and SW1 (\citealt{Korsun2006,Korsun2008}; \citealt{Ivanova2016} and this work). New results obtained by \citet{Rubin}, who investigated and detected the molecular nitrogen content in the nucleus of comet Churyumov-Gerasimenko using a mass spectrometer on board ROSINA probe "Rosetta”, are also in agreement with that statement. According to modern understanding, the distant comets are members of the Kuiper Belt and Oort Cloud and therefore they should contain a pristine material from which the Solar System bodies were created. Abundance of this molecule may give us an important information about ice condensation in protosolar nebula and delivering of the volatile elements to the terrestrial planets.\\
In this work we investigate the comet SW1. SW1 is a distant comet and is regarded to be a comet of the centaurs class. The study of centaur is still hampered by limited physical data. Therefore, every new observation and research of centaurs is important.\\

\section{Observation and reduction}

Spectrophotometric observations were made at the SOAR 4.1 m telescope in Cerro Pach\'{o}n - Chile during  August 12, 2016. The comet was observed at the heliocentric distance of 5.9 AU. The Goodman imaging/spectrograph was used with a 600 l/mm grid, which provides for the spectroscopic mode a reciprocal dispersion of 0.065 nm/pixel and, using a 1.68 arcsec slit width, a spectral element resolution of 0.73 nm. The SOAR Goodman spectrograph blue camera features one 4096 x 4096 pixel Fairchild CCD and a 7.2 arcmin in diameter field of view in the imaging mode. The seeing was stable during the night, with a mean value of 0.8 arcsec FWHM. The cometary spectra were acquired with 10 exposures of 1200 s each, which were co-added to increase the final signal-to-noise ratio. The spectrophotometric standard star LTT 9491 from \citet*{Hamuy1992, Hamuy1994} was observed with a long slit of 3 arcsec width, allowing a more precise flux calibration. Cu-Ar arcs were taken immediately after each 3 exposures of cometary spectra in order to perform wavelength calibration. Data reduction was performed using the IRAF package, following the standard procedure for CCD reduction, i.e. correction of bias and flat-field. The spectral images  were extracted and calibrated in wavelength and flux. Atmospheric extinction was corrected through mean coefficients derived for the CTIO observatory. Summarized information about the observations for SW1 is presented in Table \ref{tabl:tab1}.

\begin{table}[b]
%	\centering
	\caption{Log of the observations of comet 29P/Schwassmann-Wachmann 1}
	\label{tabl:tab1}
	\begin{minipage}{15cm}
	\begin{tabular}{lccccccc}
		\hline \hline
%	 &  &  & &  &  &  \\
	Date of observ. & Exptime, & r, & $\Delta$, & $\alpha$,\footnote{Phase angle.} &  Type & Info\footnote{Spectral resolution and filter used.} \\
	(UT) & (s) & (AU) & (AU) & ($^\circ$) &  &  \\
	\hline
		\multirow{5}{*}{2016 Aug. 12} & 1200 & \multirow{5}{*}{5.9} & \multirow{5}{*}{5.0} & \multirow{5}{*}{5.1} & Spec. & 7.28\,\AA \\
		& 300 &                   &                   &                   & \multirow{4}{*}{Phot} & B \\
		& 200 &                   &                   &                   &                    & V \\
		& 100 &                   &                   &                   &                    & R \\
		& 100 &                   &                   &                   &                    & I \\
		\hline
	\end{tabular}
\end{minipage}
\end{table}

\section{The spectroscopic data analysis}

As we know the observed spectrum of a comet is a combination of the emission spectrum of the coma gas and the reflection spectrum of the dust coma.
\begin{equation} 
\label{eq:6}
F_{comet}=F_{gas} + F_{dust},
\end{equation}
The dust spectrum F$_{dust}$ can be written as
\begin{equation}
F_{dust} = C(\lambda) · F_{solar}
\end{equation}

\begin{figure}[tp]
	\centering
	\includegraphics[width=1.0\textwidth]{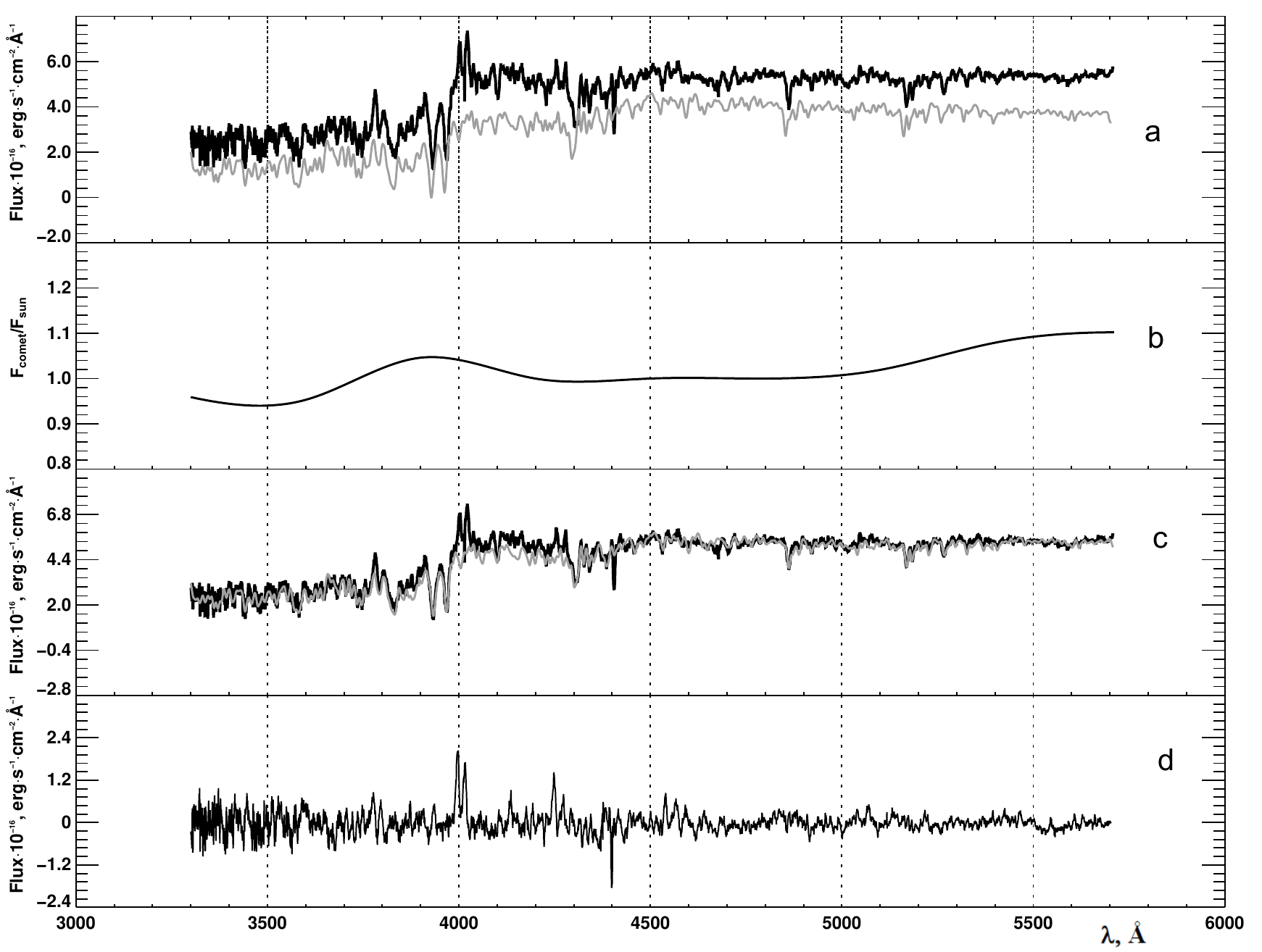}
	\caption{The results of processing of the spectrum of comet 29P/Schwassmann-Wachmann 1: (a) the energy distribution in the comet's spectrum (black line) and the shifted normalized spectrum of the Sun (gray line); (b) the normalized spectral dependence of the reflectivity of dust; (c) the spectrum of the comet (black line) and the calculated comet's continuum (gray line); (d) the emission component in the comet's spectrum.}
	\label{fig:fig1}
\end{figure}

F$_{solar}$ is the solar spectrum and C($\lambda$) is a function describing the reflectivity and scattering of the dust. In order to extract the gas emission spectrum F$_{gas}$ from the observed cometary spectrum F$_{comet}$ the function C($\lambda$) has to be determined.
There are two ways to obtain the solar spectrum F$_{solar}$. The best way would be to observe the solar spectrum directly with the same instrumental setup as the comet. Also a practical solution is to observe a solar analog star. These stars are of the same spectral class as our Sun and have therefore a nearly identical spectrum.If the solar spectrum has not been observed in the same night, the solar spectral atlas by \citet{Kurucz1984} or \citet{Neckel1984} can be used. We used the solar spectral atlas by \citet{Kurucz1984}. That solar spectrum has very high resolution. For this reason it was convolved with a Gaussian profile to decrease the resolution and normalized to the flux of the comet around 5000~\AA. The convolved solar spectrum is compared with the cometary spectrum in Fig. \ref{fig:fig1}a. Solar absorption bands have to be identical in width and depth.\\

The approach chosen to obtain C($\lambda$) is to approximate the function by a polynomial C($\lambda$) or to interpolate by spline or to use the median filter with wide window after dividing the cometary spectrum by the solar one. We used solar absorption lines in comet and solar spectra as repers and spline approximation. The result is displayed in Fig. \ref{fig:fig1}b. The spectrum C($\lambda$)$\cdot$F$_{solar}$ together with the cometary spectrum is shown in Fig. \ref{fig:fig1}c. This one is then subtracted from the cometary spectrum (Fig. \ref{fig:fig1}d). Qualitative comparison curves on the panels in Fig. \ref{fig:fig1} indicates the presence of weak cometary molecular emissions and some differences on the flux distribution between the spectra.\\
The  dust  color  indicates  trends  in  the wavelength  dependence  of  the  light  scattered  by  the  dust. Traditionally  the  color  of  cometary  dust  was determined through  measurements  of  the  comet  magnitude \textbf{m}  in  two different continuum filters.  This  color  was  a  unitless  characteristic  expressed  as the  logarithm  of  the  ratio  of  intensities  in  two  filters.  Although  this  definition  is  still  used,  spectrophotometry  of comets resulted in the definition of color as the spectral gradient of reflectivity, usually measured in $\%$ per 1000~\AA ~with an indication of the range of wavelength it was measured in.\\
From Fig. \ref{fig:fig1}b, one can see that there is a nonlinear increase with the wavelength of the scattering efficiency in blue and red regions. \\
The spectral reflectance is S($\lambda$) = F$_c(\lambda)$/F$_s(\lambda)$. Here,  F$_c(\lambda)$ is the cometary continuum, and F$_s(\lambda)$ is a scaled spectrum of the Sun. The normalized reflection ability S$^\prime$ can be described as in \citet{Jewitt1986}: 

\begin{equation} 
\label{eq:6}
S^\prime(\lambda_1,\lambda_2)=\frac{\frac{dS}{d\lambda}}{S_{mean}},
\end{equation}

where $\frac{dS}{d\lambda}$ is the rate of change of the reflectivity with respect to wavelength in the region from  $\lambda_1$ to $\lambda_2$ and S$_{mean}$ is the mean reflectivity in the observed wavelength range: 

\begin{equation} 
\label{eq:7}
S_{mean}=N^{-1} \sum S_i(\lambda).
\end{equation}

\begin{figure}[tp]
\centering
\includegraphics[width=1.0\linewidth]{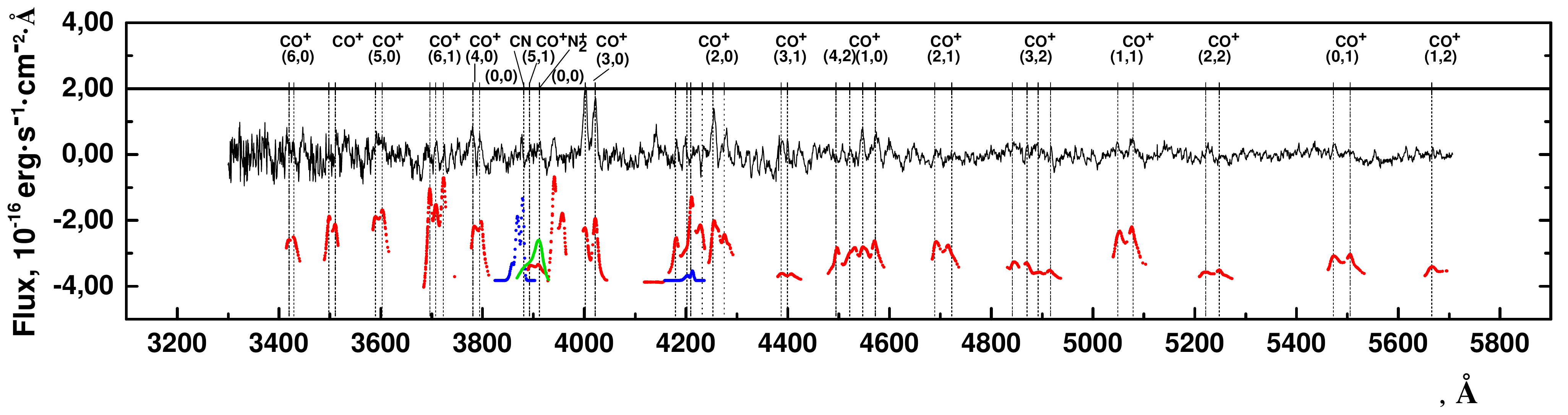}
\caption{Molecular emissions identified in the observed spectrum of comet SW1. Calculated spectra of the identified species are displayed at the bottom of the figure.}
\label{fig:fig2}
\end{figure}

In \citet{Ivanova2016} we determined the S$^\prime$($\lambda_1, \lambda_2$) for the most typical spectral regions that are defined by the effective wavelengths of the cometary filters used for the continuum registration [BC (4430\,\AA), GC (5260\,\AA) and RC (6840\,\AA)], see \citep{Schleicher2004} as 11.4\,$\pm$\,2.3\% per 10$^3$\,\AA~for the range 4430\,-\,5260\,\AA, 17.9\,$\pm$\,5.6\,\% per 10$^3$\,\AA~for the range 5260\,-\,6840\,\AA~and 14.8\,$\pm$\,4.8\% per 10$^3$\,\AA~for the range 4430\,-\,6840\,\AA. The values obtained for the normalized spectral gradient for ranges 4430\,-\,5260\,\AA~and 5260\,-\,6840\,\AA~ are comparable within the error bars to the redder wavelength region. This result does not allow an unambiguous conclusion about the normalized spectral gradient behavior with increasing wavelength.\\
Now we observed comet in spectral region 3300\,-\,5700\,\AA, therefore unfortunately we cannot use RC. But because we used the SOAR Goodman spectrograph blue camera, now we add UC (3448\,\AA) measurements. So we obtained the following results for the normalized spectral gradient: 7.80\,$\pm$\,0.02\% per 10$^3$\,\AA~for the range 3448\,-\,4450\,\AA, 6.04\,$\pm$\,0.003\,\% per 10$^3$\,\AA~for the range 4450\,-\,5260\,\AA~and 3.87\,$\pm$\,0.02\% per 10$^3$\,\AA~for the range 3448\,-\,5260\,\AA.\\
Like \citet{Ivanova2016} we made identification in the spectra of comet 29P. The SOAR Goodman spectrograph blue camera enabled us to investigate more shorter wavelengths of the spectrum, which is usually highly noisy. The linear spectrum of the comet is shown in Fig. \ref{fig:fig2}. An identification of the spectra details was made by means of comparison with laboratory and calculated molecular spectra in the same spectral region.\\

The strongest features which extend along the whole observed spectral window are the comet-tail bands of CO$^+$ \citep{Arpigny1964}. Like in our previous investigation of \citet{Ivanova2016} the lines (2,0), (3,0), (2,0), (1,0), (5,1), (3,1), (2,1), (4,2), (3,2), (0,0), and (1, 1) of the vibrational transitions of CO$^+$ (A$^2\Pi$-X$^2\Sigma$) band system are clearly seen in Fig. \ref{fig:fig2}. Additionally we have identified (6,0), (5,0) the vibrational transitions of CO$^+$ (A$^2\Pi$-X$^2\Sigma$) band system too. Two weak bands, (0, 1) and (1, 2), which belong to the (B$^2\Sigma$-A$^2\Pi$) system (Baldet \& Johnson) of the CO$^+$ ion, are detected as well. The N$_2^+$(0, 0) band of the (B$^2\Sigma$-X$^2\Sigma$) electronic system is shown too.\\
Also the [N$_2^+$]/[CO$^+$] ratio is estimated by us like in  \citet{Ivanova2016}. We used integrated intensities of the CO$^+$(2, 0) and N$_2^+$(0, 0) bands and the excitation factors of 7.0$\times$10$^{-2}$ photons$\cdot$s$^{-1} \cdot$mol$^{-1}$ for the N$_2^+$(0, 0) band \citep{Lutz1993} and 3.55$\times$10$^{-3}$ photons$\cdot$s$^{-1} \cdot$mol$^{-1}$ for the CO$^+$(2, 0) band \citep{Lutz1993, Magnani1986}. If only (2, 0) band column density of CO$^+$ is used, then [N$_2^+$]/[CO$^+$] should be equal to 0.01.This is the upper limit because  CO$^+$(5,1) bands show double peaks, and the second peak coincides with N$_2^+$ band. In \citet{Ivanova2016} we estimated the available contamination of N$_2^+$ by CO$^+$(5,1). This contamination is only 19\%. 

\section{Discussions}

\begin{figure}[tp]
	\centering
	\includegraphics[width=1.0\linewidth]{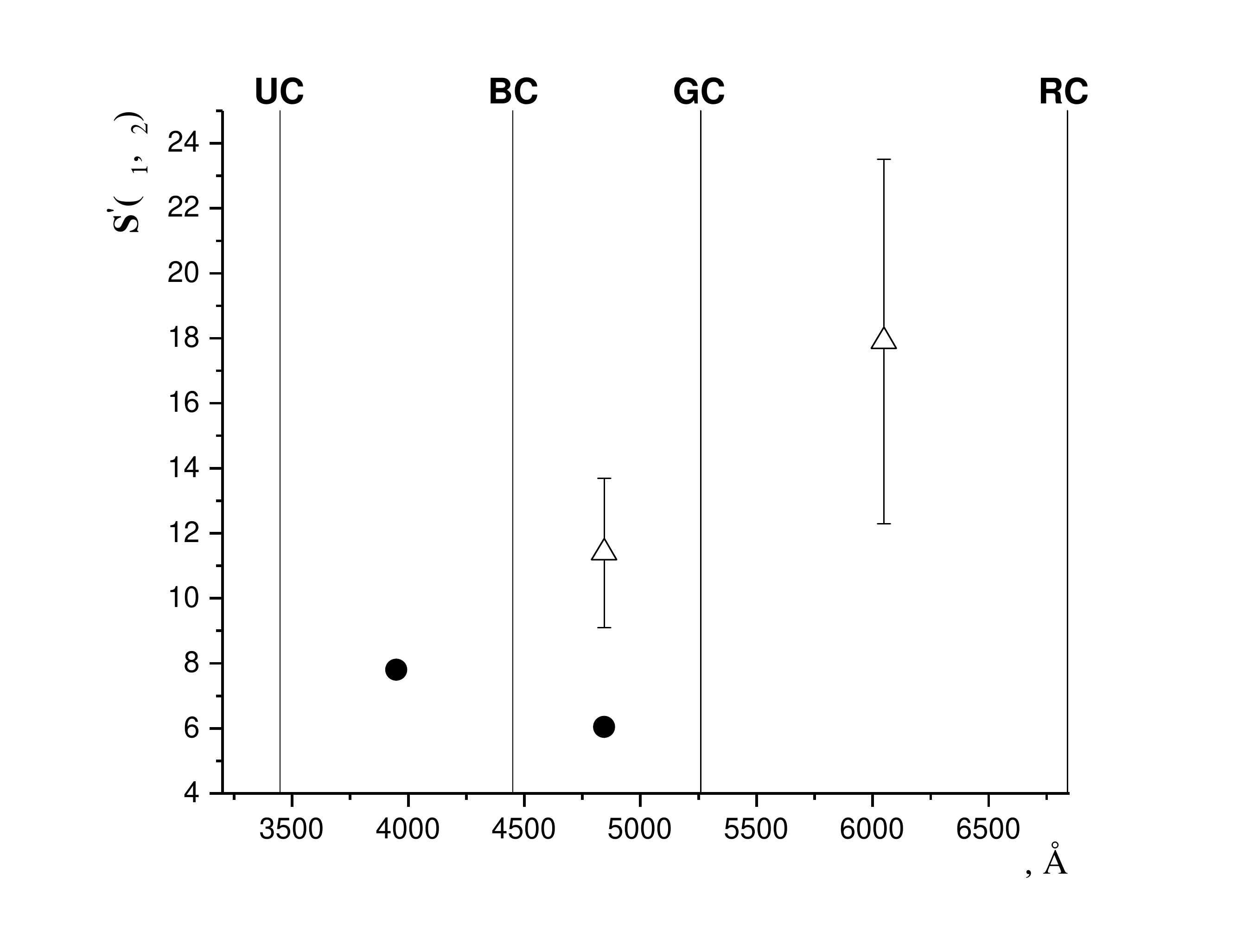}
	\caption{Behavior of the spectral gradient for comet SW1. $\bullet$ - this work; $\triangle$ - from \citet{Ivanova2016}}
	\label{fig:fig3}
\end{figure}

Our previous result \citep{Ivanova2016} does not allow an unambiguous conclusion about the normalized spectral gradient behavior with increasing wavelength. Now we can state that in the range 3448\,-\,5260\,\AA,\ the spectral gradient decreases (see Fig. \ref{fig:fig3}). 

\begin{figure}[tp]
	\centering
	\includegraphics[width=1.0\linewidth]{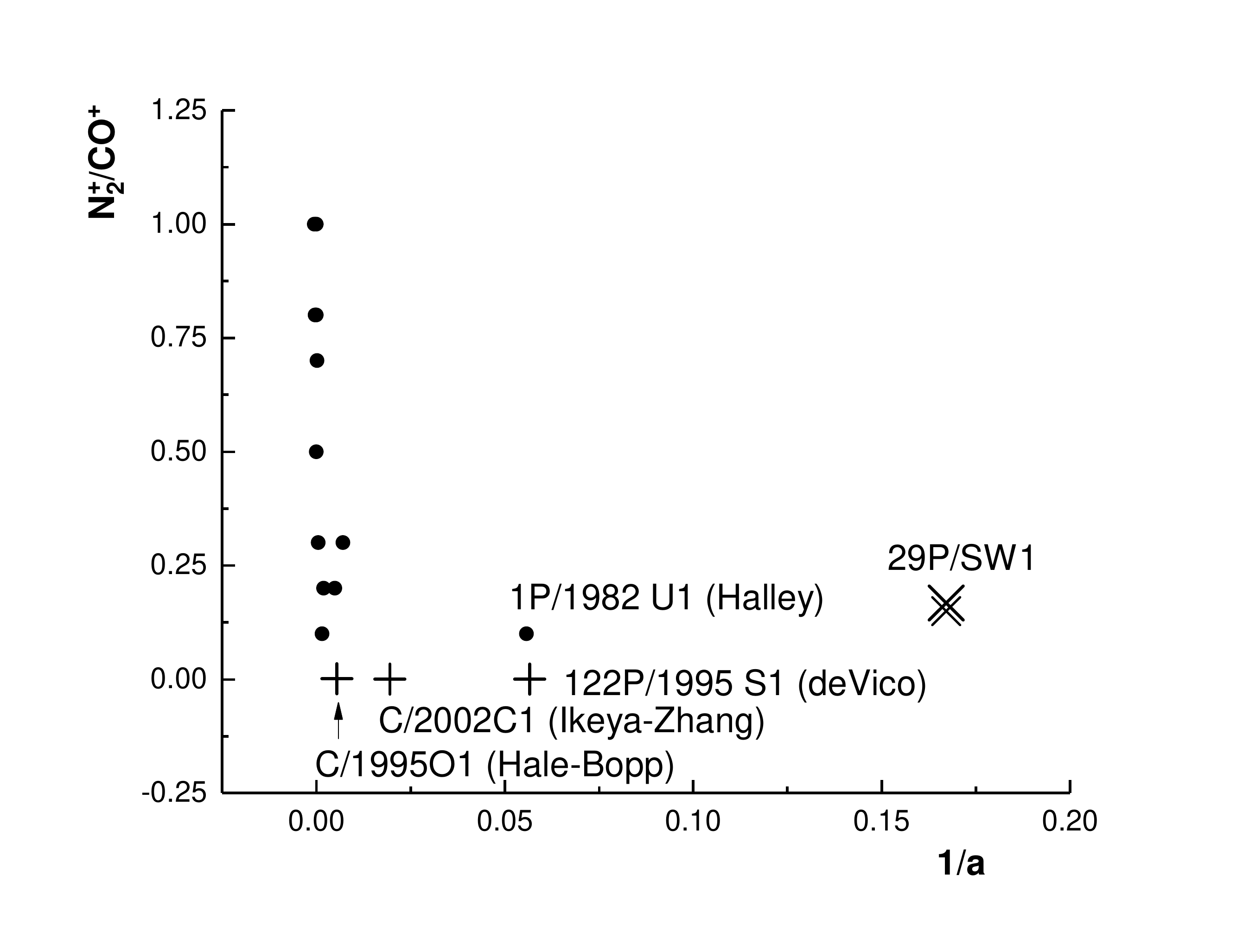}
	\caption{Estimates of N$_2^+$(0,0)/CO$^+$(4,0) for different comets. a - is semi-major axis; $\times$ - 29P/SW1 (this work and \citealt{Ivanova2016}; we converted the intensity ratios by using the relationship I(4, 0) = 0.6$\times$I(2, 0), where I(4, 0) is the intensity of the (4,0) band, I(2, 0) is the intensity of the (2,0) band, and the factor is taken from Table 4 of \citealt{Magnani1986}); $\bullet$ - from table 3 of \citealt{Cochran2000}; $+$ - from \citealt{Cochran2002} and \citealt{Cochran2000}  }
	\label{fig:fig4}
\end{figure}

Usually the reddening slope decreases towards the near-infrared  \citep*{Jewitt1986, Kolokolova2004}. But some local variations of colors have been observed in comae. The color depends on the size distribution of the grains and aggregates and on their refractive indices, mainly for grains larger than the wavelength. For instance, the jets of the Hale-Bopp comet were less red than background. It is possible that this effect was caused by the dust particle size less than one micron \citep{Furusho1999}.Comet C/1999 S4 (LINEAR) showed bluer color while the small dust grain were detected just after beginning of its disruption \citep{Hadamcik2003}. Small-size dust grains (water ice crystals, for instance) can be responsible for such a color. The rather transparent dust grains (water ice crystals, some silicates, or unaltered organic molecules) can be responsible for bluer color of the comet \citep*{Kiselev2004, Hadamcik2009, Zubko2011, Zubko2012, Hadamcik2014}.\\
 For five Jupiter-family comets, the average value is 19\% per 10$^3$\,\AA \citep{Hadamcik2009}. We have calculated \citep{Ivanova2016} the average value of the normalized spectral gradient for Centaurs (in Table 5 from \citet{Peixinho2015}). The value of the normalized spectral gradient for the Centaurs is 21.3\,$\pm$\,1.4 per 10$^3$\,\AA. We found that the mean value of the normalized spectral gradient for SW1 is 5.9\,$\pm$\,0.03\% per 10$^3$\,\AA~  and 14.7\,$\pm$\,4.2\% per 10$^3$\,\AA. The observed change in the color of the comet SW1 (Fig.\ref{fig:fig1}) can be related to its degree of activity in the observed period. \\
Knowledge of the nitrogen content of comets is important for an understanding of conditions in the early solar nebula. \citet*{Lewis} showed that conditions in the early solar nebula were such that the dominant equilibrium species of carbon, oxygen and nitrogen should be N$_2$, CO, and H$_2$O.\\
It is expected that the most abundant species in the protoplanetary nebula (together with water ice) were CO and CO$_2$ ices \citep{Meech2004}. The transitions of the CO$^+$ molecule identified in this work confirm this conclusion. According to \citet*{Capria2000a,Capria2000b} volatiles species can exist in the comet nuclei in the form of ices, as well as in the form of gaseous inclusions in amorphous water ice cells. CO sublimation begins at about 25\,K (at large distances), while CO gas release from water ice cells begins at temperatures more than 100 K \citep{Prialnik}, i.e. at  distances closer to the Sun.\\
N$_2$ molecule is rather inertious one, and therefore it is suitable to study nitrogen chemistry of comets, although  observation of N$_2$ bands is a difficult task. As a result N$_2$ was detected only in some comets. For example, \citet*{Cochran2000} report about non-detection of N$_2$ and clear detection of CO. They derived upper limits of N$_2$/CO of 3.0$\cdot$10$^{-4}$ and 0.7 to 1$\cdot$10$^{-4}$ for deVico and Hale-Bopp, respectively. The N$_2$/CO ratio is almost equal to the N$_2^+$/CO$^+$ ratio. These upper limits, compared to the NH$_3$ and CO abundances relative to water measured in comets, correspond to a very low abundance of N$_2$ in cometary ice. In comet Hale-Bopp, the N$_2$/NH$_3$ ratio is less than 0.001. Comparing with laboratory experiments on the deposition of various gases along with H$_2$O amorphous ice (Bar Nun et al., 1988), Cochran et al. (2000) concluded that N$_2$/CO is strongly depleted in these two comets. On the other hand, the N$_2$ depletion does not appear so great in other comets in which N$_2$ was possibly detected (see Table 3 from \citealt{Cochran2000}). \citet*{Cochran2000} discussed the results obtained in Comets deVico and Hale-Bopp and presented various arguments or speculations explaining the nitrogen depletion. They concluded that either a mechanism must be found to deplete the N$_2$ abundance in cometary ices after the formation of comets, or we have to understand how cometary icy grains depleted in N$_2$ may originate from a presolar cloud in which N$_2$ is currently estimated, according to the most chemical models of the interstellar medium, to be more abundant than NH$_3$. However, \citet{Charnley} showed that it is possible that much of the available nitrogen in the presolar cloud was in the form of atomic nitrogen N$_2$, with a significant contribution in NH$_3$ ice, and little contribution in N$_2$. This circumstance can explain apparent deficiency of N$_2$ seen in comets. \citet{Cochran2002} used the high-resolution spectra to search for the N$_2^+$ bands but without success. Therefore, an upper limit of 5.4$\cdot$10$^{-4}$ was set. To explain the strong depletion of N$_2$ one might suppose that the reason is the following: this molecule reluctantly form clathrate hydrates, in contrast with such molecule as CO. Last results obtained from the ROSINA mass spectrometer onboard  of ROSETTA also detected molecular nitrogen in the nucleus of comet 67P/Churyumov-Gerasimenko.  Possible mechanisms of ionization of the parent CO in cometary coma at large heliocentric distance are not fully understood, also.
We detect both N$_2^+$ and CO$^+$. We compare our estimate of N$_2^+$/CO$^+$ ratio with estimates by other authors and show the results in Fig. \ref{fig:fig4}.

%All obtained results make it possible to speak about of formation scenario of the Solar System. \\

\section{Conclusions}
	1. Spectral dependence of the light scattering by the cometary dust is obtained from the spectral observations of the comet SW1. The mean value of the normalized spectral gradient is 5.9\,$\pm$\,0.03\% per 10$^3$\,\AA.\\
	2. We detected numerous lines of CO$^+$ as well as the  N$_2^+$(0, 0) line of the (B$^2\Sigma$-X$^2\Sigma$) system, suggesting that the comet was formed in a low temperature (about 25\,K) environment. \\
	3. Additionally we have identified (6,0), (5,0) of the vibrational transitions of CO$^+$ (A$^2\Pi$-X$^2\Sigma$) band system also.\\
	4. The value of [N$_2^+$]/[CO$^+$] is equal to 0.01 for comet SW1.\\

\section*{Acknowledgements}
This research is based on observations obtained at the Southern Astrophysical Research (SOAR) telescope, which is a joint project of the Minist$\acute{e}$rio da Ci$\hat{e}$ncia, Tecnologia e Inova\c c$\tilde{a}$o (MCTI) da Rep$\acute{u}$blica Federativa do Brasil, the U.S. National Optical Astronomy Observatory (NOAO), the University of North Carolina at Chapel Hill (UNC), and Michigan State University (MSU). O. Ivanova thanks the SASPRO Programme, the People Programme (Marie Curie Actions) European Union's Seventh Framework Programme under REA grant agreement No. 609427, and the Slovak Academy of Sciences (grant Vega 2/0032/0014). I. Luk'yanyk research is supported by the project 16BF023-02 of the Taras Shevchenko National University of Kyiv and the Slovak Academy of Sciences (grant Vega 2/0032/0014).





\end{document}